\documentclass[12pt,titlepage]{article}
\usepackage[dvips]{graphicx}

\begin{document}

\title{Confinement of spinless particles by Coulomb potentials in
two-dimensional space-time}
\date{}
\author{Antonio S. de Castro \\
\\
UNESP - Campus de Guaratinguet\'{a}\\
Departamento de F\'{\i}sica e Qu\'{\i}mica\\
Caixa Postal 205\\
12516-410 Guaratinguet\'{a} SP - Brasil\\
\\
Electronic mail: castro@feg.unesp.br}
\maketitle

\begin{abstract}
The problem of confinement of spinless particles in 1+1 dimensions is
approached with a linear potential by considering a mixing of Lorentz vector
and scalar couplings. Analytical bound-states solutions are obtained when
the scalar coupling is of sufficient intensity compared to the vector
coupling.
\end{abstract}

\section{Introduction}

The potential generated by a point charge, the Coulomb potential, depends on
the dimensionality of \ space and in a 1+1 dimension it is linear (see, e.g.
\cite{eft}). As the time component of a Lorentz vector, the 1+1 dimensional
Coulomb potential is linear and so it provides a constant electric field
always pointing to, or from, the point charge. This problem is related to
the confinement of fermions in the Schwinger and in the massive Schwinger
models \cite{col1}-\cite{col2} and in the Thirring-Schwinger model \cite{fro}%
. It is frustrating that, due to the tunneling effect (Klein\'{}s paradox),
there are no bound states for this kind of potential regardless of the
strength of the potential \cite{cap}-\cite{gal}. The linear potential,
considered as a Lorentz scalar, is also related to the quarkonium model in
one-plus-one dimensions \cite{hoo}-\cite{kog}. Recently it was incorrectly
concluded that even in this case there is just one bound state \cite{bha}.
Later, the proper solutions for this last problem were found \cite{cas}-\cite%
{hil}. However, it is well known from the quarkonium phenomenology in the
real 3+1 dimensional world that the best fit for meson spectroscopy is found
for a convenient mixture of vector and scalar potentials put by hand in the
equations (see, e.g., \cite{luc}). The same can be said about the treatment
of the nuclear phenomena describing the influence of the nuclear medium on
the nucleons \cite{ser}-\cite{guo}. The mixed vector-scalar potential has
also been analyzed with the Dirac equation in 1+1 dimensions for a linear
potential \cite{cas2} as well as for a general potential which goes to
infinity as $|x|\rightarrow \infty $ \cite{ntd}. In both of those last
references it has been concluded that there is confinement if the scalar
coupling is of sufficient intensity compared to the vector coupling.

Since the pure vector Coulomb potential also frustrates the existence of
bound-state solutions in the Klein-Gordon (KG) equation \cite{gal}, the
present work proposes to consider a general mixing of vector and scalar
Lorentz structures. Such as in the case of the Dirac equation, this sort of
mixing shows to be a powerful tool to obtain a deeper insight about the
nature of the KG equation and its solutions. The problem is mapped into an
exactly solvable Sturm-Liouville problem of a Schr\"{o}dinger-like equation
with an effective harmonic oscillator potential. In two very special
circumstantes the effective potential becomes a Coulomb potential and a sort
of phase transition shows up resulting in a spectrum only consisting
exclusively of either particle-energy levels or antiparticle-energy levels.

\section{The Klein-Gordon equation with mixed vector-scalar potentials in a
1+1 dimension}

In the presence of vector and scalar potentials the 1+1 dimensional
time-independent KG equation for a spinless particle of rest mass $m$ reads

\begin{equation}
-\hbar ^{2}c^{2}\,\frac{d^{2}\psi }{dx^{2}}+\left( mc^{2}+V_{s}\right)
^{2}\psi =\left( E-V_{v}\right) ^{2}\psi  \label{1b}
\end{equation}

\noindent where $E$ is the energy of the particle, $c$ is the velocity of
light and $\hbar $ is the Planck constant. The vector and scalar potentials
are given by $V_{v}$ and $V_{s}$, respectively. The subscripts for the terms
of potential denote their properties under a Lorentz transformation: $v$ for
the time component of the 2-vector potential and $s$ for the scalar term. It
is worth to note that the KG equation is covariant under $x\rightarrow -x$
if $V_{v}(x)$ and $V_{s}(x)$ remain the same. Also note that $\psi $ remains
invariant under the simultaneous transformations $E\rightarrow -E$ and $%
V_{v}\rightarrow -V_{v}$. Furthermore, for $V_{v}=0$, the case of a pure
scalar potential, the negative- and positive-energy levels are disposed
symmetrically about $E=0$.

The KG equation can also be written as
\begin{equation}
H_{eff}\psi =-\frac{\hbar ^{2}}{2m}\,\psi ^{\prime \prime }+V_{eff}\,\psi
=E_{eff}\,\psi  \label{1c}
\end{equation}

\noindent where
\begin{equation}
E_{eff}=\frac{E^{2}-m^{2}c^{4}}{2mc^{2}},\quad V_{eff}=\frac{%
V_{s}^{2}-V_{v}^{2}}{2mc^{2}}+V_{s}+\frac{E}{mc^{2}}\,V_{v}  \label{1d}
\end{equation}

\noindent From this one can see that for potentials which tend to $\pm
\infty $ as $|x|\rightarrow \infty $ the KG equation furnishes a purely
discrete spectrum for $|V_{s}|>|V_{v}|$ , or for $|V_{s}|=|V_{v}|$ and $%
V_{s}+V_{v}E/(mc^{2})>0$. On the other hand, if the potentials vanish as $%
|x|\rightarrow \infty $ the continuum spectrum is omnipresent but the
necessary conditions for the existence of a discrete spectrum is not an easy
task for general functional forms. The boundary conditions on the
eigenfunctions come into existence by demanding that the effective
Hamiltonian given (\ref{1c}) is Hermitian, viz.

\begin{equation}
\int_{a}^{b}dx\;\psi _{n}^{*}\left( H_{eff}\psi _{n^{^{\prime }}}\right)
=\int_{a}^{b}dx\;\left( H_{eff}\psi _{n}\right) ^{*}\psi _{n^{^{\prime }}}
\label{22-1}
\end{equation}

\noindent where $\psi _{n}$ is an eigenfunction corresponding to an
effective eigenvalue $\left( E_{eff}\right) _{n}$ and $\left( a,b\right) $
is the interval under consideration. In passing, note that a necessary
consequence of Eq. (\ref{22-1}) is that the eigenfunctions corresponding to
distinct effective eigenvalues are orthogonal. It can be shown that (\ref%
{22-1}) is equivalent to

\begin{equation}
\left[ \psi _{n}^{*}\frac{d\psi _{n^{^{\prime }}}}{dx}-\frac{d\psi _{n}^{*}}{%
dx}\psi _{n^{^{\prime }}}\right] _{x=a}^{x=b}=0  \label{22-2}
\end{equation}
\noindent In the nonrelativistic approximation (potential energies small
compared to $mc^{2}$ and $E\simeq mc^{2}$) Eq. (\ref{1b}) becomes

\begin{equation}
\left( -\frac{\hbar ^{2}}{2m}\frac{d^{2}}{dx^{2}}+V_{v}+V_{s}\right) \psi
=\left( E-mc^{2}\right) \psi  \label{1e}
\end{equation}

\noindent so that $\psi $ obeys the Schr\"{o}dinger equation with binding
energy equal to $E-mc^{2}$ without distinguishing the contributions of
vector and scalar potentials.

It is remarkable that the KG equation with a scalar potential, or a vector
potential contaminated with some scalar coupling, is not invariant under $%
V\rightarrow V+const.$, this is so because only the vector potential couples
to the positive-energies in the same way it couples to the negative-ones,
whereas the scalar potential couples to the mass of the particle. Therefore,
if there is any scalar coupling the absolute values of the energy will have
physical significance and the freedom to choose a zero-energy will be lost.
It is well known that a confining potential in the nonrelativistic approach
is not confining in the relativistic approach when it is considered as a
Lorentz vector. It is surprising that relativistic confining potentials may
result in nonconfinement in the nonrelativistic approach. This last
phenomenon is a consequence of the fact that vector and scalar potentials
couple differently in the KG equation whereas there is no such distinction
among them in the Schr\"{o}dinger equation. This observation permit us to
conclude that even a ``repulsive'' potential can be a confining potential.
The case $V_{v}=-V_{s}$ presents bounded solutions in the relativistic
approach, although it reduces to the free-particle problem in the
nonrelativistic limit. The attractive vector potential for a particle is, of
course, repulsive for its corresponding antiparticle, and vice versa.
However, the attractive (repulsive) scalar potential for particles is also
attractive (repulsive) for antiparticles. For $V_{v}=V_{s}$ and an
attractive vector potential for particles, the scalar potential is
counterbalanced by the vector potential for antiparticles as long as the
scalar potential is attractive and the vector potential is repulsive. As a
consequence there is no bounded solution for antiparticles. For $V_{v}=0$
and a pure scalar attractive potential, one finds energy levels for
particles and antiparticles arranged symmetrically about $E=0$. For $%
V_{v}=-V_{s}$ and a repulsive vector potential for particles, the scalar and
the vector potentials are attractive for antiparticles but their effects are
counterbalanced for particles. Thus, recurring to this simple standpoint one
can anticipate in the mind that there is no bound-state solution for
particles in this last case of mixing. For short, when $V_{v}=\pm V_{s}$ the
spectrum only consists of energy levels either for particles or for
antiparticles.

\section{The mixed vector-scalar Coulomb potential}

Now let us focus our attention on scalar and vector potentials in the form
\begin{equation}
V_{s}=g_{s}|x|,\quad V_{v}=g_{v}|x|  \label{12}
\end{equation}
\noindent where the coupling constants, $g_{s}$ and $g_{v}$, are real
parameters. In this case the second equation of (\ref{1d}) transmutes into
\begin{equation}
V_{eff}=\frac{1}{2}\,Ax^{2}+B|x|  \label{12b}
\end{equation}
where
\begin{equation}
A=\frac{g_{s}^{2}-g_{v}^{2}}{mc^{2}},\quad B=g_{s}+\frac{E}{mc^{2}}\,g_{v}
\label{14}
\end{equation}

\noindent Therefore, one has to search for bounded solutions in an effective
shifted harmonic oscillator potential for $g_{s}^{2}\neq g_{v}^{2}$, or
Coulomb potential for $g_{s}^{2}=g_{v}^{2}$. The KG eigenvalues are obtained
by inserting the effective eigenvalues into the first equation of (\ref{1d}%
). Since the effective potential is even under $x\rightarrow -x$, the KG
eigenfunction can be expressed as a function of definite parity. Thus, we
can concentrate our attention on the positive half-line and impose boundary
conditions on $\psi $ at $x=0$ and $x=+\infty $. From (\ref{22-2}) one can
see that in addition to $\psi \left( \infty \right) =0$, the boundary
conditions at the origin can be met in two distinct ways: odd functions obey
the Neumann condition ($d\psi /dx|_{x=0}=0$) whereas even functions obey the
Dirichlet condition ($\psi \left( 0\right) =0$) .

Now we move to consider a quantitative treatment of our problem by
considering the two distinct classes of effective potentials.

\subsection{The effective shifted harmonic oscillator potential ($%
g_{s}^{2}\neq g_{v}^{2}$)}

For this class, the existence of bound-state solutions requires $%
|g_{s}|>|g_{v}|$. Included into this class is the case of a pure scalar
coupling, but the case of a pure vector, though, is naturally excluded. If $%
g_{s}<-g_{v}$ the theory is essentially relativistic. Let us define
\begin{equation}
y=|x|+B/A  \label{99}
\end{equation}
so that (\ref{1c})-(\ref{1d}) transmute into
\begin{equation}
-\frac{\hbar ^{2}}{2m}\frac{d^{2}\psi }{dy^{2}}+\frac{1}{2}\,Ay^{2}\psi
=\left( E_{eff}+\frac{B^{2}}{2A}\right) \psi  \label{100}
\end{equation}

\noindent whose square-integrable solution satisfying the boundary
conditions at $x=0$ is given by

\begin{eqnarray}
\psi (y) &=&N_{n}\exp \left( -y^{2}/2\right) H_{n}\left( y\right) ,\quad
n=0,1,2,\ldots  \nonumber \\
&&  \label{101} \\
E_{eff} &=&\left( n+\frac{1}{2}\right) \hbar \sqrt{\frac{A}{m}}-\frac{B^{2}}{%
2A}  \nonumber
\end{eqnarray}

\noindent where $N_{n}$ is a normalization constant and $H_{n}\left(
y\right) $ is a Hermite polynomial. It follows that the KG eigenenergies are
the solutions of a second-degree algebraic equation:

\begin{equation}
E=-mc^{2}\frac{g_{v}}{g_{s}}\pm \frac{\left( g_{s}^{2}-g_{v}^{2}\right)
^{3/4}}{g_{s}}\sqrt{\left( 2n+1\right) \hbar c}  \label{102}
\end{equation}

\noindent There is an infinite sequence of allowed KG eigenenergies with
alternate parities to each sign in (\ref{102}). This result clearly shows
that $g_{v}\rightarrow -g_{v}$ makes $E\rightarrow -E$ without interfering
in $\psi $, as stated earlier. The change $g_{s}\rightarrow -g_{s}$ also
makes $E\rightarrow -E$ but $\psi $ suffers a diverse sort of translation
along the $x$-axis because $B$ also changes sign. Eq. (\ref{102}) also shows
that the KG eigenenergies are symmetrically disposed about $E=0$ in the case
of a pure scalar coupling and tends to be so as $g_{s}/|g_{v}|\rightarrow
\infty $.

\begin{figure}[th]
\begin{center}
\includegraphics[width=9cm, angle=270]{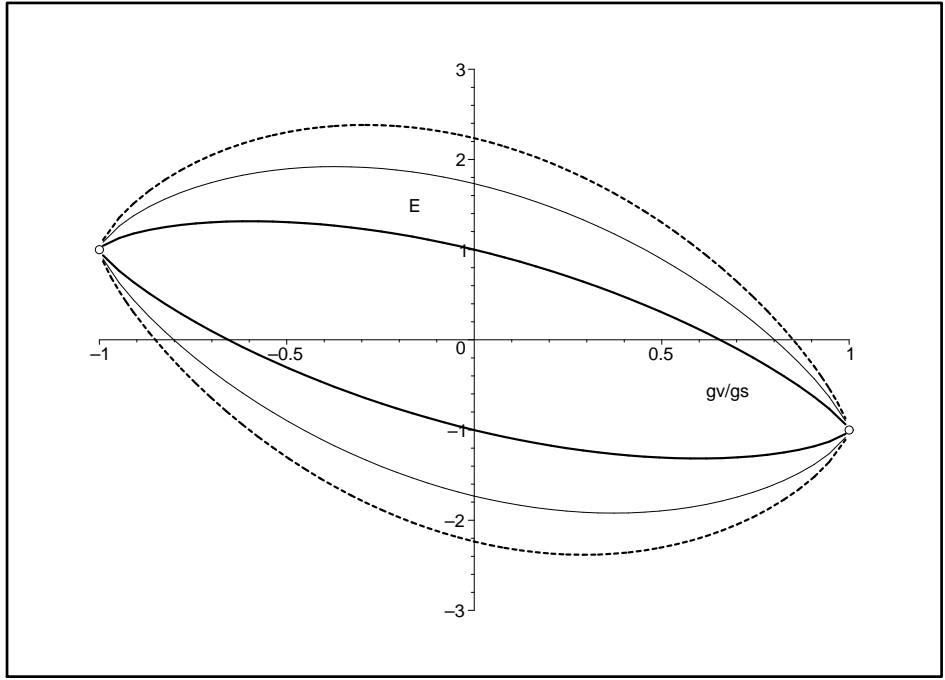}
\end{center}
\par
\vspace*{-0.1cm} \caption{KG eigenenergies for the three lowest
energy levels as a function of $g_{v}/g_{s}$ for $g_{s}=1$. The full
thick line stands for $n=0$, the full thin line for $n=1$ and the
dashed line for $n=2$ ($m=\hbar=c=1$). } \label{Fig1}
\end{figure}

\begin{figure}[th]
\begin{center}
\includegraphics[width=9cm, angle=270]{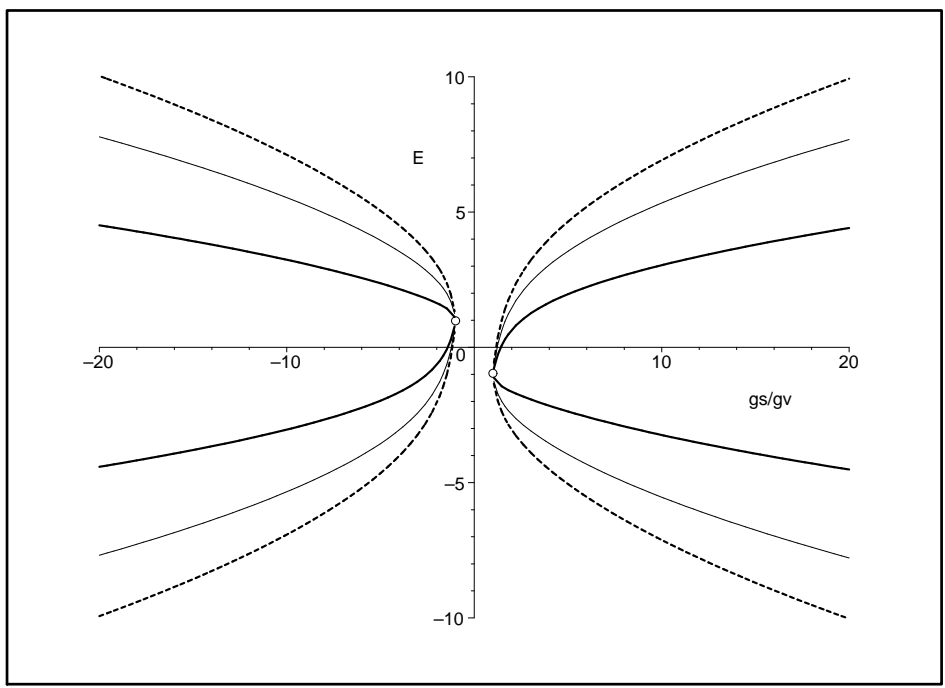}
\end{center}
\par
\vspace*{-0.1cm}
\caption{The same as in Fig. 1 as a function of $g_{s}/g_{v}$ for $g_{v}=1$.}
\label{Fig2}
\end{figure}

Fig. \ref{Fig1} (Fig. \ref{Fig2}) illustrates the three lowest energy levels
for this class of effective potential as a function of $g_{v}/g_{s}$ ($%
g_{s}/g_{v}$). These figures show that both sorts of energy levels, for
particles and antiparticles, are always present. It is evident from Fig. \ref%
{Fig1} that the bounded solutions for particles (antiparticles) are
restricted to $E>-mc^{2}$ ($E<+mc^{2}$). On the other hand, from Fig. \ref%
{Fig2} one can see that the bounded solutions for particles (antiparticles)
are restricted to $E>-mc^{2}$ ($E<-mc^{2}$) for $g_{s}>0$. Despite the
particle and antiparticle energy levels share the same energy in the
spectral gap between $-mc^{2}$ and $+mc^{2}$ for the case $g_{v}/|g_{s}|<1$,
there is no crossing of levels.

\subsection{The effective Coulomb potential ($g_{s}^{2}=g_{v}^{2}$)}

For this class of effective potential $A=0$ and the discrete spectrum arises
when $B>0$. These restrictions lead to the constraint relation
\begin{equation}
E\thinspace \mathrm{sgn}(g_{v})+mc^{2}\,\mathrm{sgn}(g_{s})>0  \label{800}
\end{equation}
Defining
\begin{eqnarray}
z &=&a|x|+b  \nonumber \\
&&  \label{20} \\
a &=&\left( \frac{2mB}{\hbar ^{2}}\right) ^{1/3},\qquad b=-E_{eff}\,\frac{a}{%
B}  \nonumber
\end{eqnarray}

\noindent Eq. (\ref{1c}) turns into the Airy differential equation

\begin{equation}
\frac{d^{2}\psi }{dz^{2}}-z\,\psi =0  \label{eq19}
\end{equation}

\noindent which has square-integrable solutions expressed in terms of the
Airy functions \cite{abr}: $\psi (z)=N\mathrm{A}_{\mathrm{i}}(z)$, where $N%
\mathcal{\ }$is a normalization constant. The boundary conditions at $x=0$
lead to the quantization conditions

\begin{eqnarray}
\mathrm{A}_{\mathrm{i}}(b) &=&0,\qquad \mathrm{for\;odd\;parity\;solutions}
\nonumber \\
&&  \label{eq19a} \\
\mathrm{A}_{\mathrm{i}}^{\prime }(b) &=&0,\qquad \mathrm{for\;even\;parity%
\;solutions}  \nonumber
\end{eqnarray}

\noindent These quantization conditions have solutions only for $b<0$ and a
number of them is listed are Table I \cite{abr}. One can see that demanding $%
b<0$ implies into an additional restriction on the KG eigenenergies: $%
|E|>mc^{2}$. Substitution of the roots of the Airy function and its first
derivative into (\ref{eq19a}) allow us to obtain the energies as the
solutions of a sixth-degree algebraic equation:
\[
E^{6}-3m^{2}c^{4}E^{4}+\left( 3m^{4}c^{6}-4|b|^{3}\hbar ^{2}g_{s}^{2}\right)
c^{2}E^{2}
\]

\begin{equation}
-8|b|^{3}\hbar ^{2}mc^{4}g_{s}^{2}\,\mathrm{sgn}(g_{s}g_{v})E-\left(
4|b|^{3}\hbar ^{2}g_{s}^{2}+m^{4}c^{6}\right) m^{2}c^{6}=0  \label{200}
\end{equation}
Since there are no explicit solutions to this algebraic equation in terms of
radicals, we satisfy ourselves verifying that the roots of Eq. (\ref{200})
always satisfy the requirement $|E|>mc^{2}$ by using the Descartes\'{} rule
of signs (henceforth DRS). The DRS states that an algebraic equation with
real coefficients $a_{k}\lambda ^{k}+...+a_{1}\lambda +a_{0}=0$ the
difference between the number of changes of signs in the sequence $%
a_{k},...,a_{1},a_{0}$ and the number of positive real roots is an even
number or zero, with a root of multiplicity $k$ counted as $k$ roots and not
counting the null coefficients (see, e.g., \cite{hand}). The verification of
the existence of solutions for $E>mc^{2}$ is made simpler if we write $%
E=mc^{2}+\varepsilon $. We get

\[
\varepsilon ^{6}+6mc^{2}\varepsilon ^{5}+12m^{2}c^{4}\varepsilon
^{4}+8m^{3}c^{6}\varepsilon ^{3}-4|b|^{3}\hbar ^{2}c^{2}g_{s}^{2}\varepsilon
^{2}
\]

\begin{equation}
-8|b|^{3}\hbar ^{2}mc^{4}g_{s}^{2}\left[ 1+\mathrm{sgn}(g_{s}g_{v})\right]
\varepsilon -8|b|^{3}\hbar ^{2}m^{2}c^{6}g_{s}^{2}\,\mathrm{sgn}%
(g_{s}g_{v})=0  \label{201}
\end{equation}

\noindent Observing the difference of signs among the coefficients of the
leading coefficient and the lowest degree \noindent it becomes clear that
there exist positive roots for $g_{v}=g_{s}$. In this particular case the
DRS assures that there exists just one solution, since there is only one
change of sign in the sequence of coefficients of (\ref{201}). For $%
g_{v}=-g_{s}$, though, the DRS implies that there exist zero or two positive
roots. It is interesting to note that this result is true whatever the
fermion mass and the coupling constants. The very same conclusion for $%
E<-mc^{2}$ can be obtained by observing that $g_{v}\rightarrow -g_{v}$ by
the change $E\rightarrow -E$. However, this is not the whole story because
the KG eigenenergies with $|E|>mc^{2}$ are not only given by the solutions
of Eq. (\ref{200}). As a matter of fact, they have also to satisfy the
constraint expressed by Eq. (\ref{800}). As a result, there are only KG
eigenenergies with $E>mc^{2}$ ($E<-mc^{2}$) for $g_{v}=|g_{s}|$ ($%
g_{v}=-|g_{s}|$). This conclusion confirms what has already been mentioned
in the last paragraph of the Sec. 2: the spectrum contains either
particle-energy levels or antiparticle-energy levels. Furthermore, it is not
difficult to see that the spectrum for the case $g_{v}=-g_{s}$ can be
obtained from that for $g_{v}=g_{s}$ by changing $E$ by $-E$.

\section{Conclusions}

In summary, we have succeed in the proposal for searching bounded solutions
of the KG equation with a mixed vector-scalar Coulomb potential. An
opportunity was given to analyze some aspects of the KG equation which have
not been approached in the literature yet.

In general, there exist bounded solutions for particles and antiparticles.
Nevertheless, for $|g_{s}|=|g_{v}|$ there are bounded solutions either for
particles or antiparticles. This is a clear manifestation of the phase
transition which occurs when $|g_{s}|=|g_{v}|$. The solutions of the KG
equation with a Coulomb potential present a continuous transition as the
ratio $g_{s}/g_{v}$ (or $g_{v}/g_{s}$) varies. However, when $%
|g_{s}|=|g_{v}|\ $the phase transition shows its face not only for the
eigenenergy but also for the eigenfunction.

Finally, we draw attention to the fact that no matter how strong the
potentials are, as far as $|g_{s}|\geq |g_{v}|$, the energy levels for
particles (antiparticles) never dive into the lower (upper) continuum. Thus
there is no room for the production of particle-antiparticle pairs. This all
means that Klein\'{}s paradox never comes to the scenario.

\vspace{1in}

\noindent{\textbf{Acknowledgments} }

This work was supported in part by means of funds provided by CNPq and
FAPESP.

\newpage

\begin{table}[tbp]
\caption{The first roots of the Airy function and its first derivative}
\label{t1}
\begin{center}
\begin{tabular}{cc}
\hline
&  \\
$Ai^{\prime }(-|b|)=0$ & $Ai(-|b|)=0$ \\
&  \\ \hline
&  \\
1.01879 & 2.33810 \\
&  \\
3.24819 & 4.08794 \\
&  \\
4.82009 & 5.52055 \\
&  \\ \hline
\end{tabular}%
\end{center}
\end{table}

\newpage

\newpage

\end{document}